\documentclass[prb,twocolumn,showpacs,amsmath,amssymb]{revtex4}
\usepackage{graphicx}
\usepackage{tabularx}
\begin{document}
%\preprint{}

\title{Numerical study of velocity statistics in steady counterflow quantum turbulence}
%\subtitle{Vortex filament Simulation with the Full Biot-Savart Law}


\author{Hiroyuki Adachi}
%\thanks{}
%\altaffiliation{Department of Physics, Osaka City University, Sumiyoshi-Ku, Osaka 558-8585, Japan}
\affiliation{Department of Physics, Osaka City University, Sumiyoshi-Ku, Osaka 558-8585, Japan}
\author{Makoto Tsubota}
\affiliation{Department of Physics, Osaka City University, Sumiyoshi-Ku, Osaka 558-8585, Japan}

\date{\today}

\begin{abstract}
 We investigate the velocity statistics by calculating the Biot--Savart velocity induced by vortex filaments in steady counterflow turbulence investigated in a previous study [Phys. Rev. B {\bf 81}, 104511 (2010)].
The probability density function (PDF) obeys a Gaussian distribution in the low-velocity region and a power-law distribution $v^{-3}$ in the high-velocity region. This transition between the two distributions occur at the velocity characterized by the mean inter-vortex distance.
Counterflow turbulence causes anisotropy of the vortex tangle, which leads to a difference in the PDF for the velocities perpendicular to and parallel to the counterflow.

\end{abstract}

%\keywords{}

\maketitle

%\section{\label{sec:intro}introduction}
Quantum turbulence, which is the disordered motion of a tangle of quantized vortices, was first observed in a counterflow experiment by Vinen.\cite{vinen} 
Since then, several properties of counterflow turbulence\cite{tough} have been revealed through numerous experimental, theoretical, and numerical studies. Although most counterflow turbulence studies have focused on statistical values, such as the vortex line density $L$ and the anisotropy of the vortex tangle, few studies have considered the PDF of the superfluid velocity, which is important in classical turbulence. There are a number of reasons for this. The first reason is that the velocity field of quantum turbulence is not measurable experimentally. Since the superfluid has no viscosity, particle image velocimetry (PIV), which is a powerful method for observing the velocity field, cannot accurately display the superfluid velocity field. Another reason is that the numerical simulation of steady counterflow turbulence has not yet been performed successfully. Schwarz performed the numerical simulation of counterflow turbulence using a vortex filament model with localized induction approximation, which neglects the interaction between vortices. As such, he was not able to obtain a statistically steady state in periodic boundary condition.\cite{schwarz88} In a previous study\cite{adachi}, we performed the numerical simulation of steady counterflow turbulence with the full Biot--Savart law, and the statistical values, such as vortex line density and the anisotropy, agree with the experimental observation. Thus, we can now numerically confirm the PDF of the superfluid velocity field in steady counterflow turbulence.

Recently, the velocity statistics of quantum turbulence has been discussed vigorously. Paoletti {\it et al.}\cite{paoletti} performed visualization of quantized vortices in a relaxation process of counterflow using solid hydrogen particles and obtained the non-classical (non-Gaussian) PDF of the particle velocity. They reported that the non-classical statistics are due to the velocity induced by the reconnection of a quantized vortex because the PDF exhibits a power-law distribution of $v^{-3}$, which is derived from the vortex velocity before or after reconnection. However, they observed the velocity of particles, which is not necessary the velocity of the superflow.
 The non-classical velocity statistics were also confirmed by White {\it et al}.\cite{white} They performed numerical simulations of quantum turbulence in a trapped Bose--Einstein condensate by calculating the Gross--Pitaevski equation to obtain the PDFs of the superflow field. The PDFs show not classical Gaussian distributions, but rather power-law distributions, which is due to the velocity field  $v=\kappa/(2\pi r)$ induced by the singular quantized vortex, where $\kappa$ is the quantum of circulation and $r$ is the distance from the core of a quantized vortex.

In the steady state of counterflow turbulence, we show that the PDF obeys non-classical distributions due to the singularity of the quantized vortex. We also reveal that the PDF depends on a line density and the anisotropy of a vortex tangle.

\begin{figure}[h]
  \begin{center}
 \scalebox{0.22}{\includegraphics{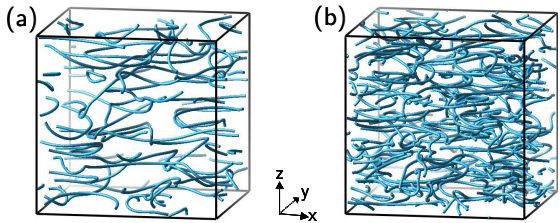}} %t19vn03
 \caption{(Color online). Configurations of the steady state of counterflow turbulence at $T=2.1 {\rm K}$. The counterflow relative velocity ${\bf v}_{sa}$ is in the direction of $-z$. (a) $v_{sa}=-0.310$ cm/s, and (b) $v_{sa}=-0.496$ cm/s.   \label{t21_steady}} 
  \end{center} 
 \end{figure}

Counterflow is the internal convection between the superfluid and the normal fluid having a relative velocity of ${\bf v}_{ns}={\bf v}_n-{\bf v}_{sa}$, where ${\bf v}_n$ and ${\bf v}_{sa}$ are the normal and superfluid velocities generated by the counterflow.
Hence, in the steady counterflow turbulence, the superflow velocity produced at a point ${\bf r}$ by a filament is given by
\begin{equation}
{\bf v}_s=\frac{\kappa}{4\pi}\int_{\cal L} \frac{({\bf s}_1-{\bf r})\times d{\bf s}_1}{|{\bf s}_1-{\bf r}|^3}+{\bf v}_{sa},
\end{equation}
where ${\bf s}_1$ refers to a point on the filament and the integration is performed along the filament.
The first term denotes the Biot--Savart velocity induced by the filament. We can obtain the superfluid velocity field by calculating Eq. (1) from the configuration of the vortices.
In the present paper, the steady counterflow turbulence is calculated in the same manner
as in Reference 4, in which we performed numerical simulations of steady counterflow turbulence using the full Biot--Savart vortex filament model, and the statistical values (vortex line density and anisotropy) agreed quantitatively with the experimental observations.
In order to investigate how the PDF depends on the vortex line density, we calculate the velocity field under two conditions, namely, ${\bf v}_{sa}=(0,0,-0.496) {\rm cm/s}$ ($L\simeq10,000 {\rm cm^{-2}}$) and ${\bf v}_{sa}=(0,0,-0.310) {\rm cm/s}$ ($L\simeq 4200{\rm cm^{-2}}$) at $T=2.1 {\rm K}$. The vortex configuration at the steady state is shown in Fig.\ref{t21_steady}. The vortex tangles are quite anisotropic in the direction of ${\bf v}_{ns}={\bf v}_n-{\bf v}_{sa}$($z$ direction). Therefore, we also investigate the influence of anisotropy on the PDF.

\begin{figure}[h]
  \begin{center}
 \scalebox{0.28}{\includegraphics{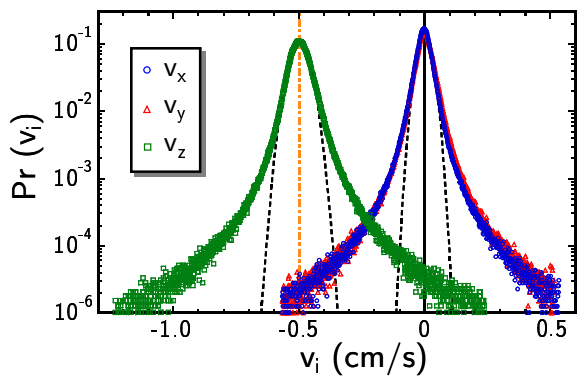}} %t19vn03
 \caption{(Color online). Probability distribution of a velocity component $v_x$, $v_y$, and $v_z$ in steady counterflow turbulence of $v_{sa}=-0.496$ cm/s at $T=2.1{\rm K}$. The vertical dot-dashed line indicates $v_{sa}$.
 \label{pdf_vx_vy_vz}} 
  \end{center} 
 \end{figure}

We calculate the superflow velocity field in steady counterflow turbulence by the spatial discretization on a $100^3$ grid. Figure \ref{pdf_vx_vy_vz} shows the PDFs of $v_x$, $v_y$, and $v_z$. Although it is known that the PDF of classical viscous turbulence is Gaussian,\cite{vincent,noullez} the PDFs of the present counterflow calculation exhibit a non-Gaussian distribution with a large tail in the high-velocity region. Since the vortex tangle of steady counterflow turbulence is isotropic in the direction perpendicular to the relative velocity ${\bf v}_{ns}$, the PDF of $v_{x}$ almost overlaps with that of $v_{y}$, with the peaks of two PDFs at $v_{x}=0$ and $v_{y}=0$.
 In contrast, since the superfluid velocity $v_{sa}=-0.496$ cm/s due to counterflow is applied in the $-z$ direction, the PDF of $v_z$ has a peak at $v_{sa}$. 

\begin{figure}[h]
  \begin{minipage}{1.0\hsize}
  \begin{center}
    \scalebox{0.22}{\includegraphics{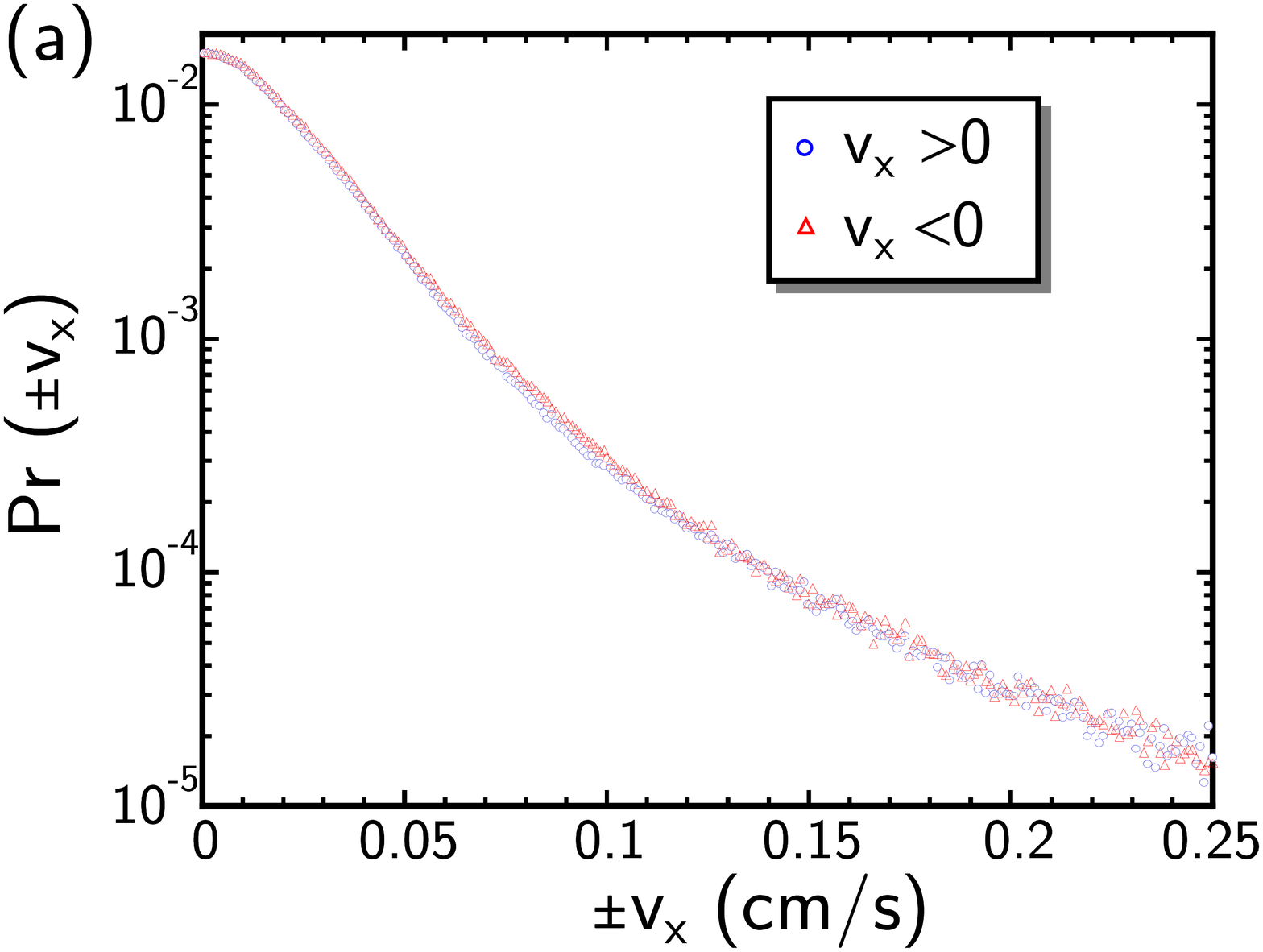}} %t19vn03
  \end{center}
  \end{minipage}
  \begin{minipage}{1.0\hsize}
  \begin{center}
    \scalebox{0.22}{\includegraphics{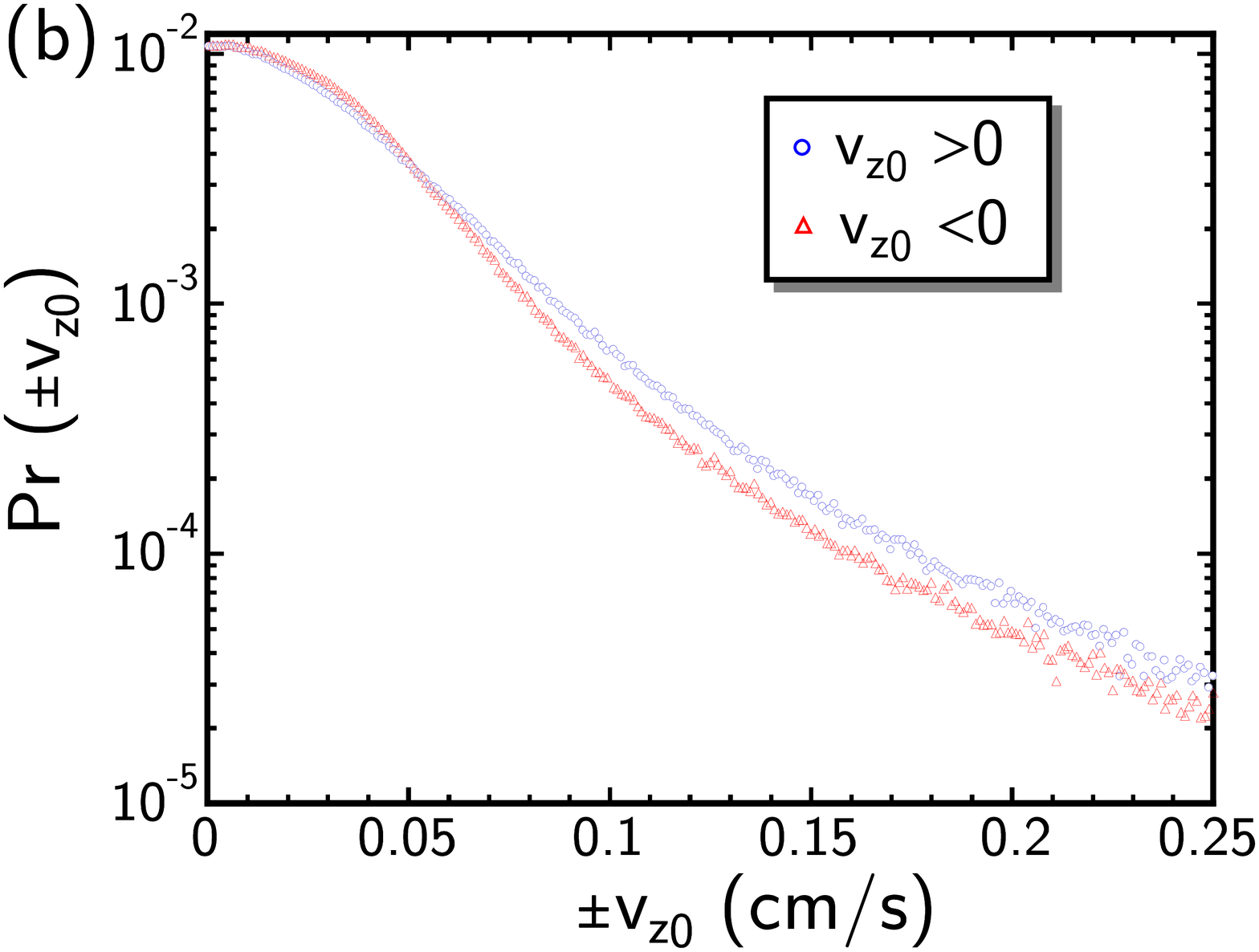}}
  \end{center}
  \end{minipage}
 \caption{(Color online). Probability distributions of (a) $v_x$ and (b) $v_z$, ($v_{sa}=-0.496$ cm/s, $T=2.1$ K) inverting the negative region around (a) $v_x=0$ and (b) $v_{z0}=0 \ {\rm cm/s}$.\label{pdf_pm}}  
 \end{figure}

Figure \ref{pdf_pm} shows the PDFs inverting the negative region around $v_{x}=v_{y}=0$ and $v_{z0}\equiv v_{z}-v_{sa}=0\  {\rm cm/s}$. Since the vortex tangle is isotropic in the direction perpendicular to ${\bf v}_{ns}$, the PDF of $v_{x}$ is symmetric to $v_x=0$. The PDF of $v_{z}$ is asymmetric to the peak $v_{z0}=0$ because the counterflow turbulence is anisotropic in ${\bf v}_{ns}$ direction. In this anisotropic state, many vortices lie in the plane perpendicular to ${\bf v}_{ns}$. In counterflow, mutual friction can expand only the vortices for which the self-induced velocity is opposite in direction to ${\bf v}_{ns}$, so that a greater number of such vortices survive, as compared to vortices for which the self-induced velocity is parallel to ${\bf v}_{ns}$. Hence, the velocity field of the counterflow vortex tangle is anisotropic in the $z$ direction.

\begin{figure}[h]
  \begin{minipage}{1.0\hsize}
  \begin{center}
 \scalebox{0.25}{\includegraphics{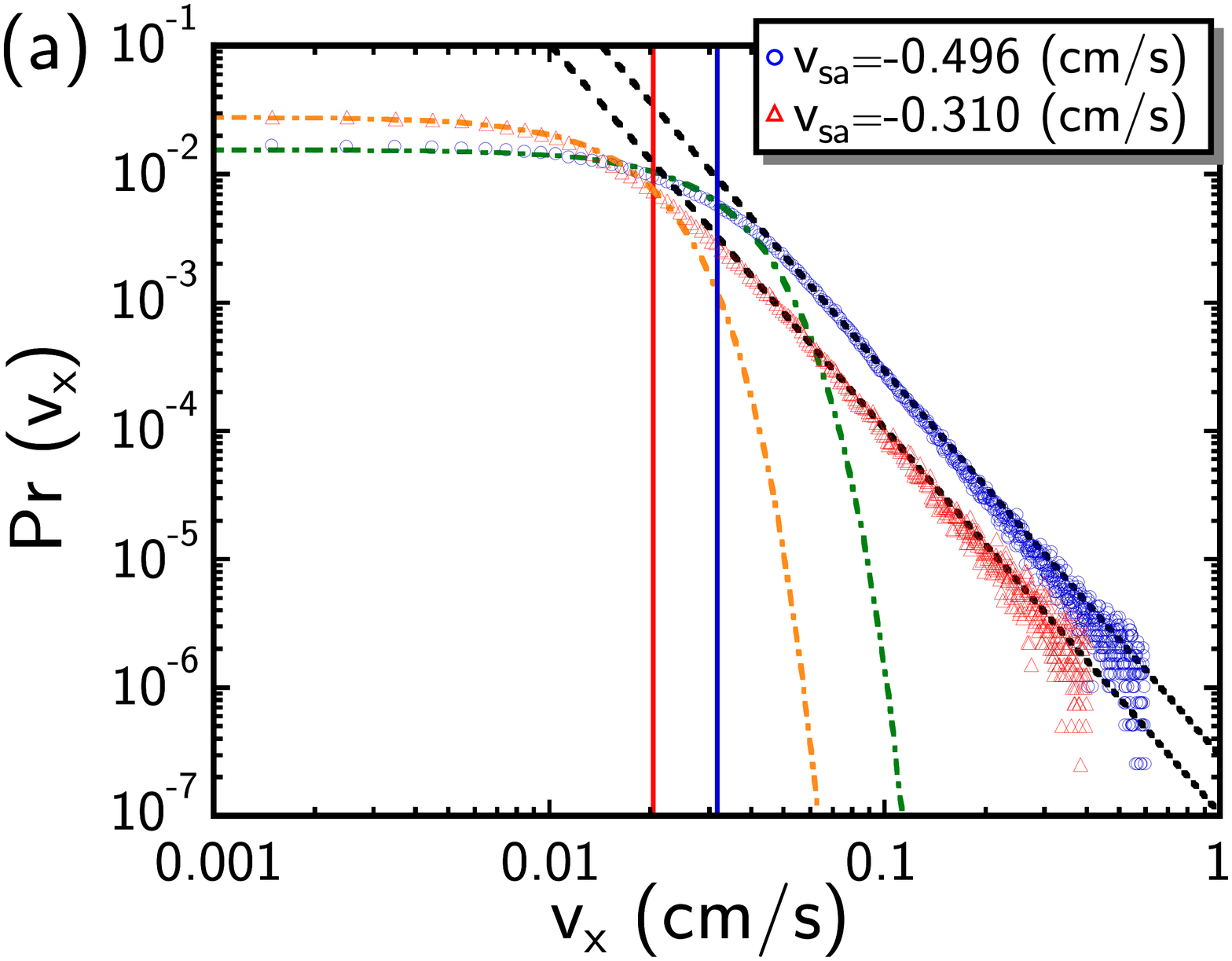}} %t19vn03
  \end{center}
  \end{minipage}
  \begin{minipage}{1.0\hsize}
  \begin{center}
 \scalebox{0.25}{\includegraphics{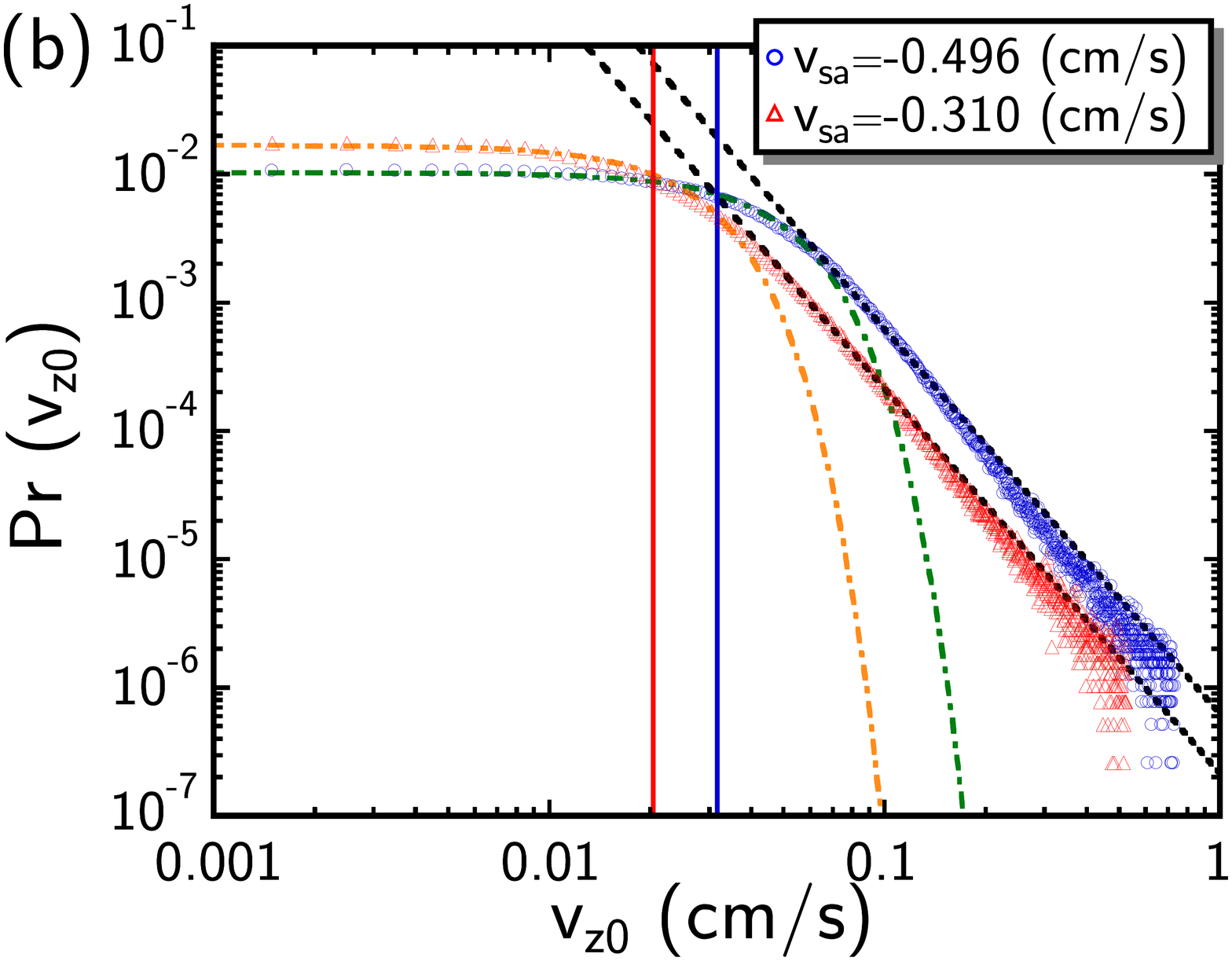}}
  \end{center}
  \end{minipage}
 \caption{(Color online). Log-log plot of the probability distributions of (a) $v_x$ and (b) $v_{z0}$ in the steady state vortex tangle calculated under $v_{sa}=-0.496\ {\rm cm/s}$ \ ($L=10,000\  {\rm cm^{-2}}$) and $v_{sa}=-0.310\ {\rm cm/s}$ \ ($L\simeq 4200\ {\rm cm^{-2}}$). The blue and red solid lines indicate the $v_t$ derived from the vortex line density $L=10000\ {\rm cm^{-2}}$ and $L=4200\ {\rm cm^{-2}}$, respectively. 
The dashed lines show $\propto v^{-3}$, and the dot-dashed lines show the Gaussian distribution.
 \label{pdf_ryoulog}}  
 \end{figure}

Figure \ref{pdf_ryoulog} shows double logarithmic plotted PDFs of the $v_x$ and $v_z$ components. Both distributions exhibit a power-law distribution of ${\rm Pr}(v_i)\propto v^{-3}_i$ ($i=x,y,z0$) in the higher-velocity region. These velocity statistics have been investigated in previous studies.\cite{min} For the single, straight vortex case, the probability of separation occurring between $r$ and $r+dr$ is $2\pi r dr$, and  the velocity scales as $1/r$, which leads to ${\rm Pr}(v)\sim {\rm Pr}(r(v))|dr/dv| \sim 1/v^3$. The PDF converges to a Gaussian distribution in the low-velocity region, probably because the vortex configuration is random in the tangle. This property of the PDF was investigated in a previous study \cite{weiss} in which the authors numerically set $N$ vortex points at random positions, and the PDF of the velocity at the vortex point was obtained. The transition of the PDF between a Gaussian distribution and a power-law distribution of $v^{-3}$ was also confirmed.

\begin{figure}[h]
  \begin{center}
 \scalebox{0.2}{\includegraphics{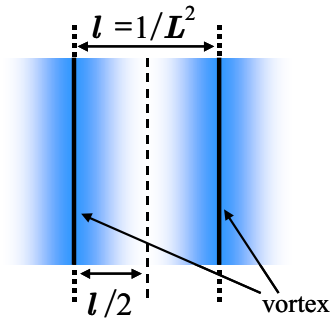}} %t19vn03
 \caption{(Color online). Schematic diagram of the velocity field induced by two parallel vortices. The color density corresponds to the magnitude of the velocity induced by the vortex, which attenuates with $1/r$. \label{two_vortex}} 
  \end{center} 
 \end{figure}

We can roughly estimate the transition velocity from the Gaussian distribution to the power-law distribution. In order to easily understand the velocity field induced by multiple vortices, we consider the simple case of two straight parallel vortices, as shown in Fig. \ref{two_vortex}. Although primarily the $1/r$ velocity appears near each vortex, in the halfway range between vortices, the velocity becomes complicated because the velocities induced by the two vortices become comparable and interfere with each other. Hence, the statistics of velocity appear to change near the midpoint between vortices. In the vortex tangle, the mean inter-vortex distance is denoted by $l=1/L^{1/2}$, and so the midpoint between the vortices is located at $l/2$. Thus, the transition velocity of the statistics should be represented by 
\begin{equation}
v_t=\frac{\kappa}{2\pi(l/2)}.
\end{equation}
In Fig. \ref{pdf_ryoulog}, the vertical solid lines refer to the estimated transition velocity $v_t$. The actual transition velocity for the power-law distribution approximately agrees with the estimated transition velocity $v_t$. The mean inter-vortex distance in the $z$ direction is different from that in the $x$ and $y$ directions due to the anisotropic vortex tangle, which leads to a difference in the actual transition velocity between the PDFs for $v_x$ and $v_z$.

In conclusion, we investigate the velocity statistics by calculating the Biot--Savart velocity induced by vortex filaments in steady counterflow turbulence investigated in a previous study.\cite{adachi} The probability density function (PDF) obeys a Gaussian distribution in the low-velocity region and obeys a power-law distribution $v^{-3}$ in the high-velocity region. 
The transition between these two distributions occurs at the velocity characterized by the mean inter-vortex distance. The anisotropy of the vortex tangle due to counterflow leads to a difference in the PDF between the velocities perpendicular to and parallel to the counterflow. In the future, we intend to study how the velocity statistics are related to other statistical quantities, such as energy spectra.\cite{kolmogorov}

 M. T. acknowledges the support of a Grant-in-Aid for Scientific Research from JSPS (Grant No. 21340104).

\end{document}